\documentstyle[prl,aps,epsf,twocolumn]{revtex}
\begin{document}

\input epsf
\draft
\twocolumn[\hsize\textwidth\columnwidth\hsize\csname
@twocolumnfalse\endcsname
\title{Fine Structure in the Off-Resonance Conductance of Small
Coulomb-Blockade Systems}
\author{J. J. Palacios$^{\ast\dagger}$,
Lerwen Liu$^{\dagger\dagger}$, and D. Yoshioka$^{\dagger\dagger}$}
\address{$^\ast$Department of Physics and Astronomy, University of Kentucky,
Lexington, KY 40506, USA.\\
$^\dagger$Department of Physics, Indiana University, Bloomington,
IN 47405, USA. \\
$^{\dagger\dagger}$ Department of Basic Science,
University of Tokyo, Komaba, Tokyo 153, JAPAN}
\date{\today}
\maketitle

\widetext
\begin{abstract}
\leftskip 2cm
\rightskip 2cm
We show how a fine, multiple-peak structure can arise in the
off-resonance, zero-bias conductance of Coulomb blockade systems.  In
order to understand how this effect comes about one must abandon the
orthodox, mean-field understanding of the Coulomb blockade phenomenon
and consider quantum fluctuations in the occupation of the single-particle
electronic levels. We illustrate such an effect with a spinless
Anderson-like model for multi-level systems and an equation-of-motion
method for calculating Green's functions that combines two simple
decoupling schemes.
\end{abstract}
\pacs{\leftskip 2cm PACS numbers: 73.23.Hk} 
\vskip2pc]

\narrowtext
Resonant tunneling through small, isolated, multi-level systems such as
a quantum dot\cite{ashoori} causes a peak in the zero-bias conductance
whenever a single-electron level coincides with the chemical potential
($\mu$) in the leads. Without Coulomb interaction a series of peaks with
separation $\Delta\epsilon$ (energy level spacing) would be observed
when, for instance, the energy levels are lowered relative to $\mu$ by
external gates. At each peak the number of electrons in the system
increases by one, filling the corresponding single-electron state.
Coulomb interactions, however, have drastic effects on the
conductance: (i) On the one hand the current can be suppressed over a
large range of gate voltages since incoming electrons may be strongly
repelled by those already present in the system.  This phenomenon is
generally known as Coulomb blockade\cite{ashoori} (CB). A mean-field
type picture\cite{beenakker} suffices to describe such an effect: Each
time an additional electron is added to a single-particle state in the
system all the other energy levels are shifted with respect to their
previous values by an amount $U$ which is related to the Coulomb
repulsion.  Therefore, if $U \gg \Delta \epsilon$, a sparse series of
conductance peaks with separation $U$ is expected instead of that with
separation $\Delta\epsilon$. (ii) A more exotic phenomenon can take
place when, at very low temperatures, the conductance between CB peaks
(off-resonance conductance) is enhanced assisted by quantum
fluctuations in the single-electron, degenerate  (usually spin-degenerate)
levels\cite{ng,meir3,kondo}.  This effect is closely related to the
Kondo effect which is well known in the literature\cite{hewsonbook}.
Below the Kondo temperature\cite{hewsonbook}, what determines which one
of the above mentioned phenomena dominates the conductance properties of
the system is the ratio $\Delta\epsilon/\Gamma$, where $\Gamma$ is the
coupling strength to the leads. For $\Delta\epsilon/\Gamma \gg 1$, the
mean-field picture is basically correct and CB physics
dominates\cite{ng}.   For $\Delta\epsilon/\Gamma \ll 1$,
fluctuations can take over and so can Kondo-type
physics\cite{ng,kondo,meir3}.

Although the CB phenomenon has been experimentally well established in
quantum dots\cite{ashoori}, to our knowledge, the latter one is yet to
be observed in such systems, in part, due to the extremely low
temperatures required. In this work we show that the presence of quantum
fluctuations in the occupation of the electronic levels can also be
playing a significant role close to the limit where CB dominates. From
our results we conclude that there is a possibility of observing an
enhancement of the off-resonance current through small quantum dots
which carries a direct fingerprint of the discrete single-particle levels.
Specifically we address the problem on how the mean-field picture breaks
down as $\Delta\epsilon/\Gamma$ crosses over from $\gg 1$ to $\gtrsim
1$. We find that in the regime $\Delta\epsilon/\Gamma \gtrsim 1$ quantum
fluctuations in the single-particle levels create "dynamical" channels
which are available for transport.  In this limit, these new channels
give rise to a multiple-peak structure in the off-resonance conductance
in addition to the smooth signature of virtual tunneling processes
which is known as elastic cotunneling\cite{cotunnel}.

We begin by considering a multi-level system connected to
left and right leads described by the Hamiltonian 
\begin{eqnarray}
H&=&\sum_{i=1}^{N_L} \epsilon_i d^\dagger_i d_i+ \sum_{k\in R,L} E_k
c^\dagger_k c_k +
\sum_{j>i=1}^{N_L} U_{ij}d^\dagger_i d_i d^\dagger_j d_j + \nonumber \\
&&\sum_{i=1}^{N_L}\sum_{k\in R,L} V_i(k)[c^\dagger_k d_i + d_i^\dagger c_k],
\end{eqnarray}
where $d_i^\dagger$ ($d_i$) are the creation (annihilation) operators
associated with the $N_L$ single-particle levels in the system
with energies $\epsilon_i$, $c_k^\dagger$ ($c_k$) are the ones for the 
levels in the left and right leads with energies {\bf $E_k$ }, and $V_i(k)$ are
the hopping  matrix elements between them.
The third term in the Hamiltonian takes care of the electronic correlations 
within the system. Such a term contains the necessary 
contributions (those of the type density-density interaction) 
to study the most fundamental aspects of transport through quantum dots.
(Additional terms might be added if one is interested in more detailed
correlation effects\cite{juanjo}, but this lies beyond the scope of
this work.)  Degeneracies like those due to the spin degree of freedom
are not considered either (a high magnetic field may be implicitly
assumed).

The conductance $g$  through the interacting system can be calculated with
the formula\cite{meir2}
\begin{eqnarray}
g=\frac{2e^2}{h}\int_{-\infty}^\infty d\omega f_{FD}'(\omega)
\Im m\{tr[\gamma_{ij}(\omega)G_{ij}(\omega)]\},
\label{g}
\end{eqnarray}
where $f_{FD}'(\omega)$ is the derivative of 
the Fermi-Dirac distribution function,
{\boldmath $\gamma$}$(\omega)$ is the hopping
 matrix defined by ${\mbox{\boldmath
$\gamma$}}^R(\omega){\mbox{\boldmath $\gamma$}}^L(\omega)/
[{\mbox{\boldmath$\gamma$}}^R(\omega)+
{\mbox{\boldmath$\gamma$}}^L(\omega)]$,  where
$\gamma_{ij}^{R(L)}(\omega)\equiv  -2\Im m[\Sigma_{ij}^{R(L)}(\omega)]=
-2\Im m\left[\lim_{\delta \rightarrow 0}
\sum_{k\in R(L)}\frac{V_i(k)V_j(k)}{\omega - E_k +i\delta}\right]$,
{\boldmath $G$}$(\omega)$ is the retarded Green's
function that must be calculated in equilibrium, and ''$tr$'' denotes the
trace over the levels of the interacting region\cite{firstnote}.

Equation-of-motion (EOM) techniques for calculating Green's functions
have been used in the past in the
context of the Anderson model\cite{eom} and, recently, also in the
context of quantum dots.\cite{meir3,meir1,pals}
According to such a technique one may write
\begin{equation}
\omega\langle\langle d_i ;d^\dagger_j\rangle\rangle =
\langle\{d_i,d^\dagger_j\}\rangle+\langle\langle[d_i,H];
d^\dagger_j\rangle\rangle,
\label{eqeom}
\end{equation}
where $ G_{ij}(\omega) \equiv \langle\langle d_i;d^\dagger_j\rangle\rangle$.
The higher-order Green's functions generated by the last term in Eq.\
(\ref{eqeom}) must be approximated at some stage of the calculation
to obtain a closed set of equations.  The simplest way to do that 
is in the Hartree-Fock approximation (HFA) where the
higher-order Green's functions generated in the first EOM cycle are
decoupled in the following way:
$\langle\langle d_in_j;d_i^\dagger\rangle\rangle \approx
\langle n_j\rangle \langle\langle d_i;d_i^\dagger\rangle\rangle$,
with $\langle n_j\rangle\equiv\langle d^\dagger_j d_j \rangle$.
In this approximation the Green's function takes the simple form   
${\mbox{\boldmath $G$}}(\omega)=[({\mbox{\boldmath $G$}}^0(\omega))^{-1}-
{\mbox{\boldmath $\Sigma$}}(\omega)]^{-1}$,
where {\boldmath $G$}$^0(\omega)$ is the diagonal Hartree-Fock 
Green's function for the isolated system 
\begin{equation}
G^0_{ii}(\omega)=\lim_{\delta \rightarrow 0}
\frac{1}{\omega-\epsilon_i-\sum_j U_{ij} \langle n_j\rangle +i\delta},
\end{equation}
and ${\mbox{\boldmath $\Sigma$}}(\omega)={\mbox{\boldmath
$\Sigma$}}^R(\omega)$+
${\mbox{\boldmath $\Sigma$}}^L(\omega)$ is the total coupling
self-energy. The Green's function projected on a level
$i$ depends on the occupation
numbers $\langle n_j \rangle$ of all the other levels
which are usually calculated self-consistently. Now,
in order to include 
dynamical processes (which will turn out to be relevant in certain limits)
in the Green's function,
one has to improve upon the HFA.  The way to do that 
in the framework of the EOM method consists in 
generating higher-order Green's functions
from $\langle\langle d_in_j ;d_i^\dagger\rangle\rangle$, namely,
$\langle\langle c_k d_j^\dagger d_i;d_i^\dagger\rangle\rangle$,
$ \langle\langle c_k^\dagger d_id_j;d_i^\dagger\rangle\rangle$,
$\langle\langle d_in_jn_l ;d_i^\dagger\rangle\rangle$, and 
$\langle\langle c_k n_j ;d_i^\dagger\rangle\rangle$.
Following Hewson and Zuckerman\cite{hewsonandzuck},
we neglect Green's functions  like
$\langle\langle c_k d_j^\dagger d_i;d_i^\dagger\rangle\rangle$
and $ \langle\langle c_k^\dagger d_id_j;d_i^\dagger\rangle\rangle$
which contain unpaired operators.  Physically, this
corresponds to considering only one-electron processes.
By generating new Green's functions from the remaining terms
and successive decouplings similar to those mentioned in the HFA,
a closed set of equations is obtained. We will refer to this approximation
as Hewson-Zuckerman approximation (HZA) from now on. 
After some lengthy algebra, and excluding off-diagonal terms,\cite{note}
we find the following expression for the retarded Green's function:
\begin{eqnarray}
G_{ii}(\omega)&=&\frac{1}{\omega-\epsilon_i-\Sigma_{ii}(\omega)}\left[
1+ \sum^{N_L}_{j=1,\ne i} \frac{U_{ij}\langle n_j \rangle}{\omega-\epsilon_i-
U_{ij}-\Sigma_{ii}(\omega)}\right. + \nonumber \\
& & +  \sum^{N_L}_{k>j=1, \ne i}
\frac{U_{ij}U_{ik}\langle n_j n_k \rangle}{\omega-
\epsilon_i-U_{ij}-U_{ik}-\Sigma_{ii}(\omega)} \times \nonumber \\
& &\left(\frac{1}{\omega-\epsilon_i-U_{ij}-\Sigma_{ii}(\omega)}+
\frac{1}{\omega-\epsilon_i-U_{ik}-\Sigma_{ii}(\omega)} \right) \nonumber \\
&&+\sum^{N_L}_{l>k>j=1, \ne i} \frac{U_{ij}U_{ik}U_{il}\langle n_j n_k n_l
\rangle}{\omega-\epsilon_i-U_{ij}-U_{ik}-U_{il}-
\Sigma_{ii}(\omega)}\times\nonumber\\
&&\left(\frac{1}{[\omega-\epsilon_i-U_{ij}-\Sigma_{ii}(\omega)]
[\omega-\epsilon_i-U_{ij}-U_{ik}-\Sigma_{ii}(\omega)]}+ \right. \nonumber \\
&&+\frac{1}{[\omega-\epsilon_i-U_{ij}-\Sigma_{ii}(\omega)]
[\omega-\epsilon_i-U_{ij}-U_{il}-\Sigma_{ii}(\omega)]}+ \nonumber \\
&&+\frac{1}{[\omega-\epsilon_i-U_{ik}-\Sigma_{ii}(\omega)]
[\omega-\epsilon_i-U_{ik}-U_{ij}-\Sigma_{ii}(\omega)]}+ \nonumber \\
&&+\frac{1}{[\omega-\epsilon_i-U_{ik}-\Sigma_{ii}(\omega)]
[\omega-\epsilon_i-U_{ik}-U_{il}-\Sigma_{ii}(\omega)]}+ \nonumber \\
&&+\frac{1}{[\omega-\epsilon_i-U_{il}-\Sigma_{ii}(\omega)]
[\omega-\epsilon_i-U_{il}-U_{ij}-\Sigma_{ii}(\omega)]}+ \nonumber \\
&&+\left.\left.\frac{1}{[\omega-\epsilon_i-U_{il}-\Sigma_{ii}(\omega)]
[\omega-\epsilon_i-U_{il}-U_{ik}-\Sigma_{ii}(\omega)]}\right) + \dots \right]
\label{g11}
\end{eqnarray}
where additional terms containing products in $U$ 
up to $U^{N_L-1}$ are present, but
are not shown here. As in the HFA, the occupancies $\langle n_j \rangle$ and
correlation functions $\langle n_j \dots n_l\rangle$ may be calculated
self-consistently. 

\begin{figure}
\epsfxsize=8cm \epsfbox{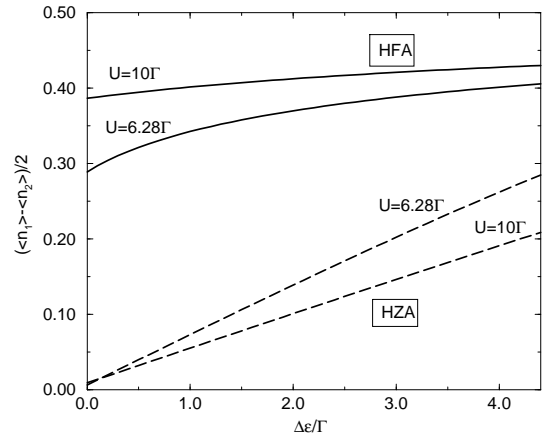}
\caption{Difference in the occupation numbers of the two levels in the
symmetric case as a function of $\Delta\epsilon/\Gamma$.
The HFA and the HZA are shown for two values of $U$. The HFA gives
the correct limit for large values of $\Delta\epsilon/\Gamma$ when
compared with, e.g., second order perturbation
theory in $U$ (see text). In the HZA, however, the
difference is too small and decreases with $U$ instead of increasing.}
\label{fig1}
\end{figure}

The HZA, as well as many other self-consistent EOM approximations beyond
the HFA,\cite{meir3,meir1,pals} presents, however, a serious drawback:
{\em It gives unphysical values for the occupation numbers}.   Figure
\ref{fig1} illustrates this shortcoming in the simplest two-level case.
As usual,\cite{hewsonbook} we take the total coupling self-energy
independent of $\omega$, equal for both levels, and purely imaginary,
$\Sigma_{jj}(\omega)=-i\Gamma$.  We plot half the difference in the
occupation numbers of  the two levels as a function of
$\Delta\epsilon/\Gamma$ when the chemical potential lies in between the
singly- and doubly-occupied states (notice that this corresponds to
plotting the magnetization for the usual symmetric Anderson
model\cite{hewsonbook,eom}). It is well known from other methods that
the fluctuations in both levels are asymptotically suppressed as one
increases $\Delta\epsilon/\Gamma$ and/or $U/\Gamma$, and that,
eventually, the HFA occupation numbers are basically
correct.\cite{ng,hewsonbook,zlatic,sorensen} For instance, second-order
perturbation theories in $U$ on top of the Hartree-Fock
solution\cite{zlatic} give $(\langle n_2 \rangle - \langle n_1
\rangle)/2\approx 0.4$ for $\Delta\epsilon/\Gamma \approx 4$ and
$U\approx 2\pi \Gamma$, which is what we obtain in the HFA. By contrast,
the self-consistent HZA gives a much smaller value. Moreover, this
value, instead of increasing with $U/\Gamma$, {\em decreases}. One way
to get around this difficulty and to obtain realistic results out of the
dynamical expression (\ref{g11}) is to avoid the fully self-consistent
procedure.  This can be effectively done by using the HFA static results
for the occupation numbers and multiple-particle correlation functions
in the expression (\ref{g11}) for the Green's function (we will
hereafter call this approximation HFA-HZA). In this way, as we approach
the isolated-system limit, $\Gamma \rightarrow 0$, the correct
occupancies are guaranteed while keeping open the possibility for
fluctuations at finite hopping. We  expect this approximation to give
reliable qualitative results in the limit $\Delta\epsilon/\Gamma \gtrsim
1$.  It cannot be valid, however, for $\Delta\epsilon/\Gamma \lesssim 1$
for several reasons: (i) The HFA breaks spontaneously the local symmetry
when $U/\Gamma >1$ (see Fig.\ \ref{fig1}); (ii) off-diagonal elements
have been ignored, and (iii) in order to reliably account for the strong
fluctuation effects, which occur in that limit, one should have gone
beyond the HZA in the EOM method\cite{eom,meir1}.

\begin{figure}
\epsfxsize=8cm \epsfbox{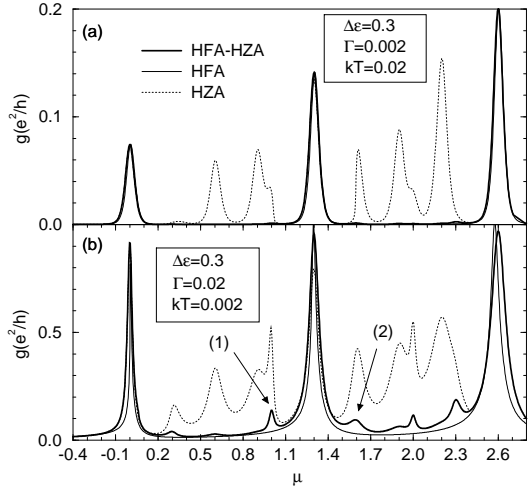}
\caption{Conductance in the HFA-HZA (thick solid line), in the HFA
(thin solid line), and in the HZA  (dotted line)
as a function of $\mu$ in a five-level system up to $ N =3$. 
We have set $\epsilon_1=0$. The other parameters are: $\Delta\epsilon =0.3$, 
$U_{ij}=1$, (a) $kT=0.02$, $\Gamma=0.002$,  (b)
$kT=0.002$, $\Gamma=0.02$
(all the magnitudes are in units of $U$). The smaller peaks in (b)
correspond to dynamical channels opened by the fluctuating occupation numbers
of the individual single-particle levels.}
\label{fig2}
\end{figure}

We now calculate the conductance of a five-level system where an
increasing coupling of the levels to the leads,
$\Sigma_{jj}(\omega)=-i(\Gamma\times j)$, simulates a realistic
situation for quantum dots. All the interaction terms $U_{ij}$ are set
to one.  Figure \ref{fig2} shows $g$ vs. $\mu$ in two different limits.
When $\Delta\epsilon/\Gamma \gg 1$ [Fig.\ \ref{fig2}(a)] both HFA and
HFA-HZA give similar results: There is a peak in the conductance
whenever $\mu \approx E(N)-E(N-1)\equiv\mu_{dot}(N)$ where $E(N)$ is the
ground state total energy of $N$ particles in the dot. The
single-particle states are successively filled and remain fully occupied
as $\mu$ moves up between the renormalized single-particle levels. As
expected,   due to the derivative of the Fermi-Dirac distribution
function in Eq.\ \ref{g}, the heights and widths of the peaks are
proportional to $1/T$ and $T$, respectively.   In contrast, the level
occupancies obtained from the HZA when off resonance [away from the
charge-degeneracy points where $\mu \approx \mu_{dot}(N)$] are
non-integer numbers and the total charge in the dot is not perfectly
quantized. (Detailed analysis will be given elsewhere.\cite{otranote})

As $\Gamma$ increases [Fig.\ \ref{fig2}(b)] the
occupation of the single-particle levels is no longer either strictly
zero or one as a function of $\mu$.  The mean-field picture rendered by
the HFA is only approximately valid and quantum-mechanical fluctuations
mediated by the interaction play their role now. In fact, within the
HFA-HZA, new peaks appear in the off-resonance conductance as a
consequence of these fluctuations.  The presence of the  peak labeled
(1) at  $\mu=1.0$ can be understood in the following way: There is a
finite probability for the electron to be in any of the levels 1, 2, 3,
4, and 5. It spends most of its time in level 1, but if it happens to be
in levels 2, 3, 4, or 5 a second, external electron can enter the system
through level 1.  Schematically these processes can be represented like
$|\circ\bullet\circ\dots\rangle \rightarrow
|\bullet\bullet\circ\dots\rangle \rightarrow
|\circ\bullet\circ\dots\rangle$, $|\circ\circ\bullet\dots\rangle
\rightarrow |\bullet\circ\bullet\dots\rangle \rightarrow
|\circ\circ\bullet\dots\rangle$, and so on, where empty and filled dots
represent empty and occupied states at a given time, respectively.
Although the cost in energy for these processes is the same,
$\epsilon_j+\epsilon_1+U_{1j}-\epsilon_j=1.0$, their likelihood
decreases with $j$.  Peak (2) at $\mu=1.6$ can be understood in a
similar way: There is a small but finite probability for the system to
have the level 2 empty even when $\mu>\mu_{dot}(2)$. This is taken
advantage of by an electron in the lead to sneak through via level 3.
Schematically: $|\bullet\circ\circ\dots\rangle \rightarrow
|\bullet\circ\bullet\dots\rangle \rightarrow
|\bullet\circ\circ\dots\rangle$.  The major peaks lie basically where
the HFA predicts and all the other minor peaks can be associated with
dynamical processes like those described above. [As can be seen, the
self-consistent HZA overestimates the importance of these type of
processes and gives rise to the spurious off-resonance structure seen
Fig.  \ref{fig2}(a).]

It is worth mentioning that the tunneling processes described above seem
to contribute to the off-resonance conductance in a way different from
the usual elastic cotunneling\cite{cotunnel} which also anticipates a
finite value of the off-resonance conductance. In such a theory, the
contribution to the conductance comes from second- or higher-order
(virtual) tunneling processes.  Virtual transport is already included in
the HFA where, in addition, divergencies close to the resonances are
automatically taken care of.  As can be seen in Fig.\ \ref{fig2}(b), the
elastic cotunneling is latent, for instance, in the slight asymmetry of
the first major resonance. In agreement with previous
work\cite{cotunnel}, the $\Delta\epsilon^2/(\Delta\epsilon+U)^2$
dependence of the off-resonance conductance can also be obtained in the
HFA . In Fig.\ \ref{fig2}(b), however, we see that the off-resonance
value in the HFA is always a lower limit of the one obtained in the
HFA-HZA. 

\begin{figure}
\epsfxsize=8cm \epsfbox{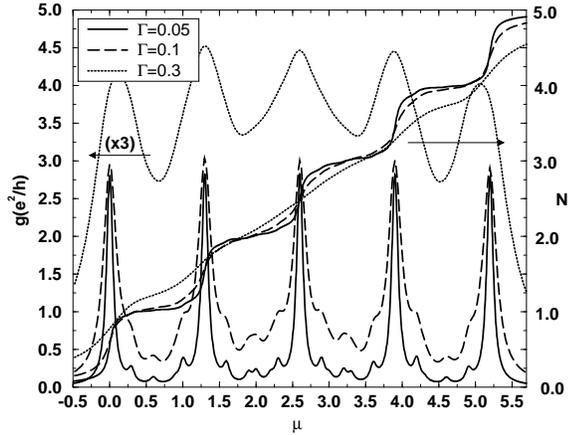}
\caption{Conductance in the HFA-HZA and total charge
as a function of $\mu$ in a five-level system for different values of
$\Gamma$ (no dependence with the single-particle
level has been considered now). 
$kT=0.005$ and all the other parameters are as in 
Fig.\ \protect{\ref{fig2}}.
As $\Gamma$ increases, the fine structure
starts disappearing as well as the overall CB
effect. One can also see how the charge quantization is lost
progressively.}
\label{fig3}
\end{figure}

Finally, we would like to stress the limits of validity of our results.
At low enough temperatures, the off-resonance, fine structure relies on
the existence of small quantum fluctuations in the occupation of the
electronic levels.  As we see from Fig.\ \ref{fig1}, such fluctuations
are suppressed when $U\gg\Delta\epsilon$ or $\Delta\epsilon \gg \Gamma$,
but they survive if $U\gtrsim\Delta\epsilon$ and $\Delta\epsilon \gtrsim
\Gamma$.  Whereas the latter condition can be easily obtained by tuning
gate voltages, the former one can only be generically found in
relatively small quantum dots (diameter typically $\lesssim 0.1\mu$m).
(Such quantum dots exist at present but their transport properties have
not been fully analyzed to date.) When $\Gamma \approx \Delta\epsilon$
one expects, not only the fine structure to disappear, but the CB
phenomenon altogether\cite{glazman} (at least for a system with an
infinite number of levels). This can be understood very easily. CB
disappears when the conductance of the insulating barriers, which keep
the dot isolated from the leads, becomes of the order $e^2/h$. This is
equivalent to saying that $\Gamma \rho_{dot} \approx 1$, where
$\rho_{dot}$ is the density of states of the dot. Since, to a first
approximation, $\rho_{dot} \approx 1/\Delta\epsilon$ we obtain
$\Gamma/\Delta\epsilon\approx 1$ as condition for the disappearance of
the CB peaks. This trend is also confirmed by our calculations in the
HFA-HZA for five levels as can be seen in Fig.\ \ref{fig3}. As the
broadening of the single-particle levels approaches $\Delta\epsilon$,
the charge quantization is lost progressively and, consequently, the
distinctive CB peaks change into a smooth oscillation as a function of
$\mu$.  However, as we discussed before, a quantitative
analysis of this regime is beyond the scope of our present work.

In conclusion, an EOM that combines two different decoupling schemes has
been used to calculate the conductance through a general multi-level
system with strong interactions. A novel multiple-peak structure is
obtained in certain limits. Whereas the main peaks
are easily understood in terms of the orthodox Coulomb Blockade theory,
quantum fluctuations must be invoked to explain the smaller peaks.
The structure created by these small peaks in the off-resonance
conductance lies on top of the contribution coming from virtual elastic 
tunneling.

We acknowledge useful discussions with H. Fertig, L. I. Glazman,
A. H. MacDonald, A. Mart\'{\i}n-Rodero, M. E.
Raikh, E. Sorensen, C. Tejedor, and H. Yi.  This work has been supported
by the National Science Foundation under grants DMR-9416902 and
DMR-9503814.  JJP acknowledges support from NATO postdoctoral research
fellowship. LL acknowledges support from the Yokoyama Scholarship
foundation in Japan.  DY and JJP thank the hospitality of Indiana 
University and Aspen Center for Physics where part of this work
was done.

\begin{references}
\bibitem{firstnote} Equation (\ref{g}) is valid only if the left and
right hopping matrices differ from each other by a constant.
In a general case a different expression for the
conductance must be used (see Ref.\ \onlinecite{meir2}).
\bibitem{note} For the two-level case an expression for the 
Green's function which includes off-diagonal terms has been also
derived. When comparing results with and without off-diagonal terms
no qualitative differences were found for the coupling strengths
we are interested in. Details will be given elsewhere 
(Ref.\ \onlinecite{otranote}).
\end{references}
\end{document}